# In-orbit calibration status of the Insight-HXMT

Xiaobo Li, Liming Song, Xufang Li, Ying Tan, Yanji Yang, et al.

# In-orbit calibration status of the Insight-HXMT

Xiaobo Li*[a], Liming Song[a], Xufang Li[a], Ying Tan[a] ,Yanji Yang[a], Mingyu Ge[a]
[a]Key Laboratory of Particle Astrophysics, Institute of High Energy Physics, Beijing, China, 100049

## ABSTRACT

As China's first X-ray astronomical satellite, Insight-HXMT (Hard X-ray Modulation Telescope) successfully launched on Jun 15, 2017. It performs timing and spectral studies of bright sources to determine their physical parameters. HXMT carries three main payloads onboard: the High Energy X-ray telescope (HE, 20-250 keV, NaI(Tl)/CsI(Na)), the Medium Energy X-ray Telescope (ME, 5-30 keV, Si-Pin) and the Low Energy X-ray telescope (LE, 1-15 keV, SCD). In orbit, we have used the radioactive sources, activated lines, the fluorescence line, and Cas A to calibrate the gain and energy resolution of the payloads. The Crab pulsar was adopted as the primary effective area calibrator and an empirical function was found to modify the simulated effective areas. The absolute timing accuracy of HXMT is about 100us from the TOA of Crab Pulsar.

**Keywords:** X-ray instrumentation, calibration, HXMT

## 1. INTRODUCTION

The Hard X-ray Modulation Telescope (HXMT) is a large X-ray astronomical satellite with a broad band in 1-250 keV. It was successfully launched on 15th June 2017 in China. It is a low earth orbit telescope with altitude of 550km and inclination of 43 degrees. In order to fulfill the requirements of the broad band spectral and variability observations, three payloads are configured onboard HXMT, which are, High Energy X–ray telescope (HE) using 18 NaI(Tl)/CsI(Na) scintillation detectors for 20-250 keV band, Medium Energy X-ray telescope (ME) using 1728 Si-PIN detectors for 5-30 keV band, and Low Energy X-ray detector (LE) using 96 SCD detectors for 1-15 keV band. The three payloads are integrated on a same supporting structure to achieve the same pointing direction, thus they can simultaneously observe the same source. They all have collimators to confine different kinds of field of view (FOV). Figure 1 shows the three main payloads onboard Insight-HXMT and their FOVs.

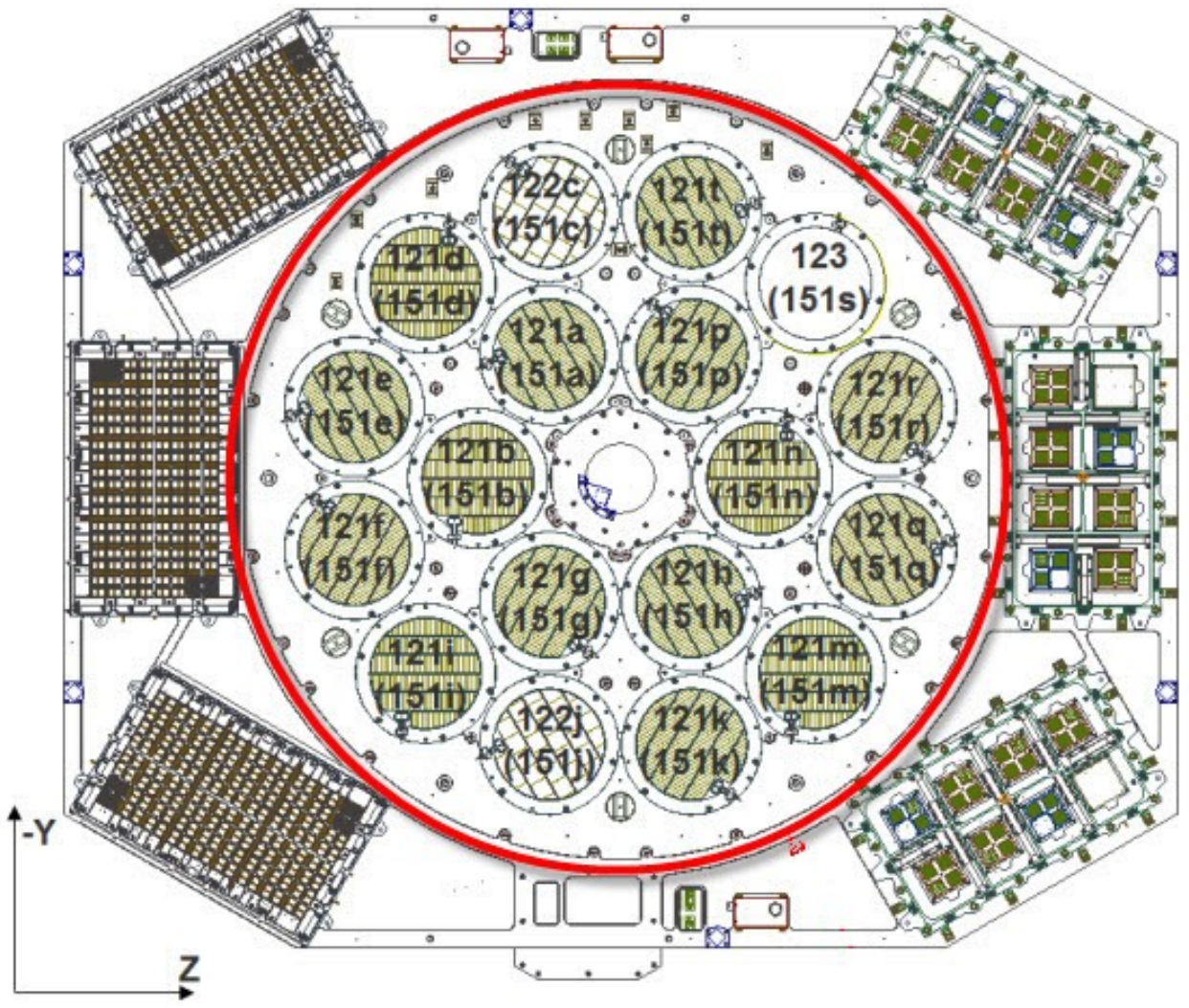

Figure 1: Three main payloads onboard Insight-HXMT. In the red circle, there are 18 NaI(Tl)/CsI(Na) scintillation detectors for HE. The left side is three ME boxes. The right side is three LE boxes. They all have different FOVs.

## 1.1 High Energy X-ray telescope

Similar to BeppoSAX/PDS and RXTE/HEXTE, HE adopts an array of NaI(Tl)/CsI(Na) PHOSWICHs as the main detectors. The diameter of each PHOSWICH is 190mm. The thickness of NaI(Tl) and CsI(Na) is about 3.5mm and 40mm. The working temperature of PHOSWICHS is actively controlled at $18\pm2$℃. The incident X-ray with most of its energy deposited in NaI(Tl) is regarded as a NaI(Tl) event. CsI(Na) is used as an active shielding detector to reject the background events from backside and events with partial energy loss in the NaI(Tl). The scintillation photons generated within the two crystals can be collected by the same photomultiplier tube(PMT). The decay time is 250ns in NaI(Tl) and 630ns in CsI(Na). Signals from the PMT(Hamamatsu R877-01) are pulse shaped to distinguish NaI(Tl) events and CsI(Na) events. The energy loss, time of arrival and the pulse width with each detected event are measured, digitized and telemetered to the ground. The CsI(Na) can be used as a gamma ray burst(GRB) monitor. The detected energy range in normal mode is about 50-800keV and is changed to 250keV-3MeV in GRB mode for CsI(Na) if the high voltage of PMT is decreased. In this paper, we only refer to the calibration of NaI(Tl) events in norm mode.

The collimators of HE define 15 narrow FOV(1.1° x 5.7° ), 2 wide FOV(5.7° x 5.7° ) and a blind FOV which was covered with 2mm tantalum. They also have different orientations with a step of 60 degree, as shown in Figure 1.

For each PHOSWICH detector, a radioactive source $^{241}$Am with an activity of 200 Bq is embedded into a plastic scintillator(BC-448M) and viewed by a separate Multi-Pixel Photon Counter. They all mounted in the collimator and used as an automatic gain control(AGC) detector as shown in the Figure 2. A coincident measurement between the AGC detector(5.5MeV alpha particle) and PHOSWICH detector(59.5keV X-ray) will be labeled as a calibration event. The calibration events are saved like norm events, just have a different flag to discriminate. The spectrum of calibration events is shown is Figure 3. In-flight performance shows that the line centroids of 59.5keV are stable to better than 0.01 channels on a one day timescale. The response of NaI(Tl) is not uniform in its large surface so the calibration events were just used as the gain control and were not suitable to calibrate the gain of NaI(Tl) detectors in-orbit.

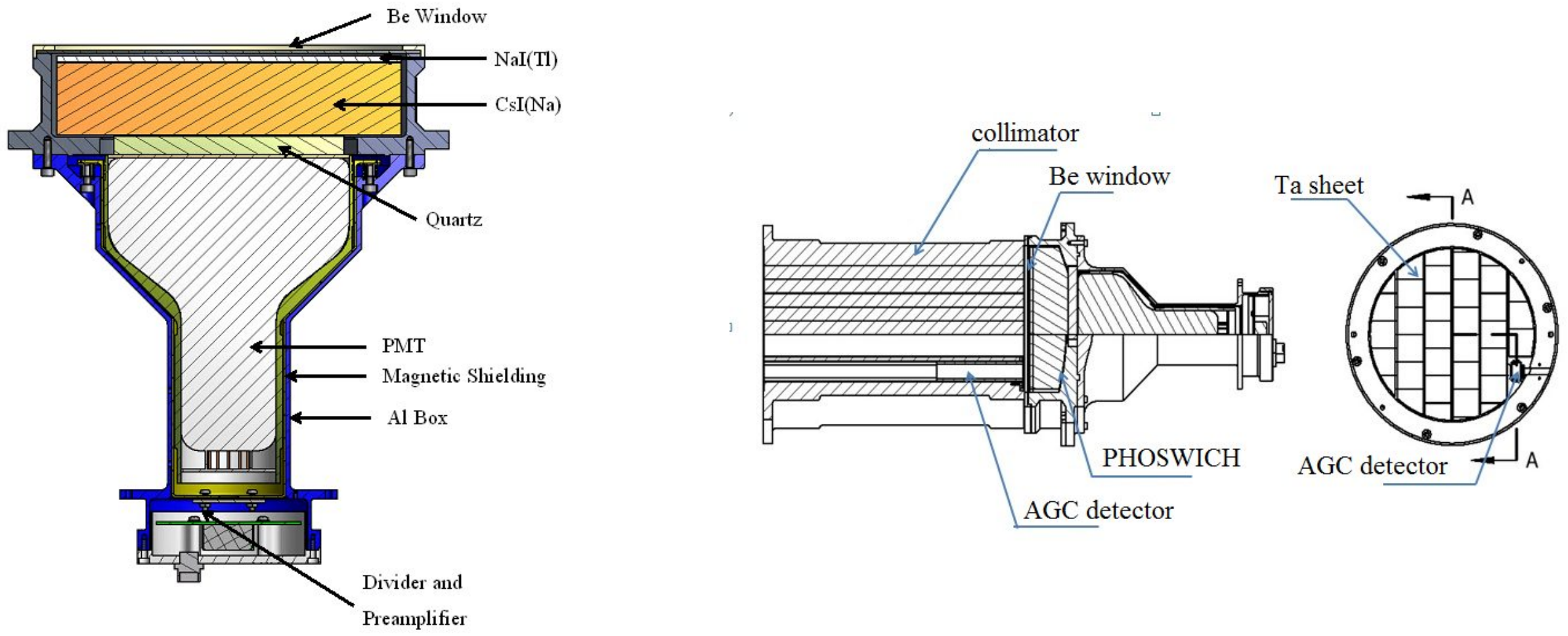

Figure 2: One detector module for HE. The AGC detector is in the later side of the PHOSWICH.

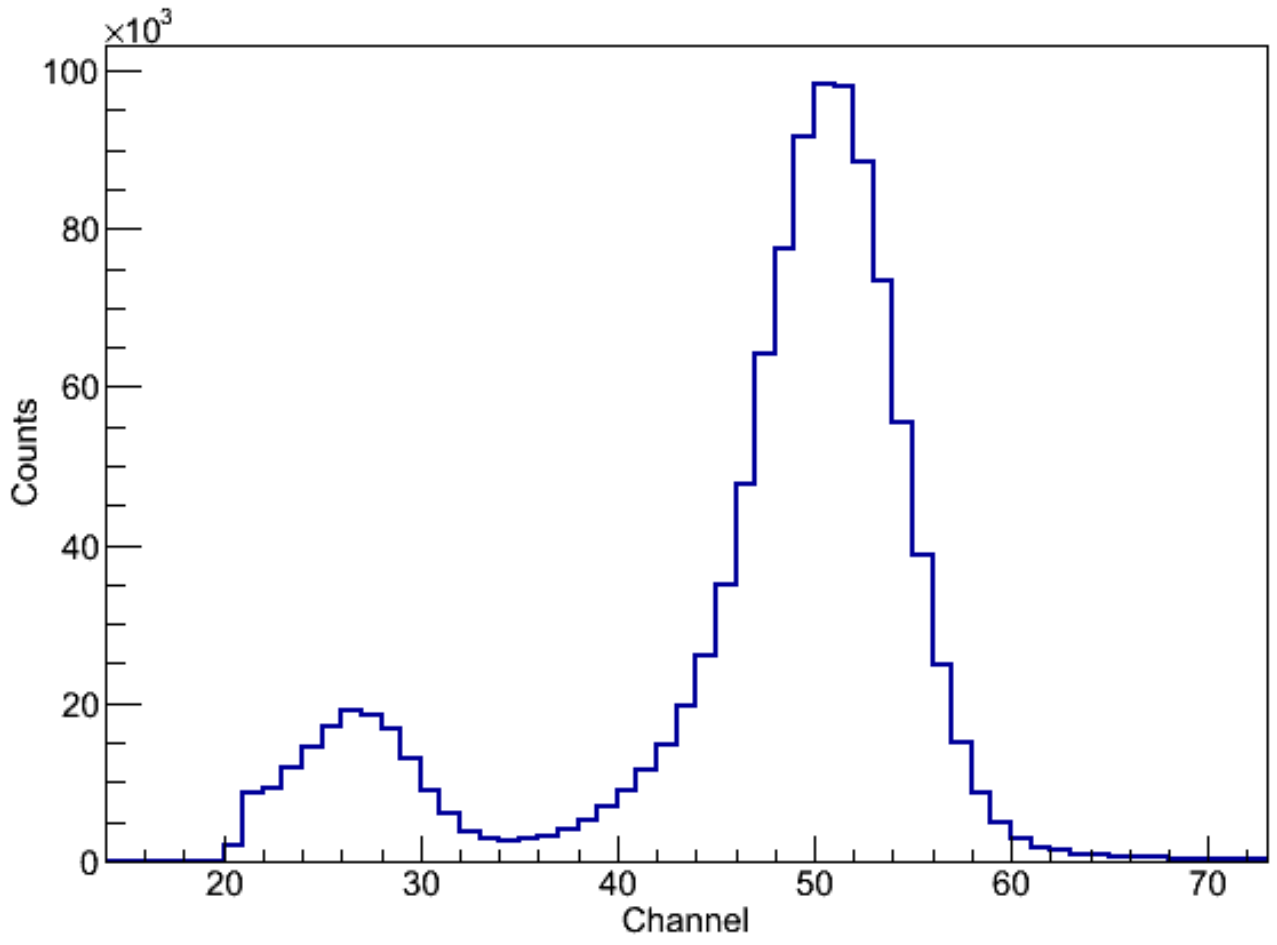

Figure 3: Calibration spectrum of $^{241}$Am. The data were accumulated over one day. The 59.5 keV line was corresponded to 50 Channel.

Besides the active shielding of CsI(Na), HE also adopts the 18 plastic scintillators(6 on the top and 12 in the later sides of PHOSWICH detectors) as the anti-coincidence detectors(ACD). Figure 4 shows examples of one top ACD and one lateral ACD.

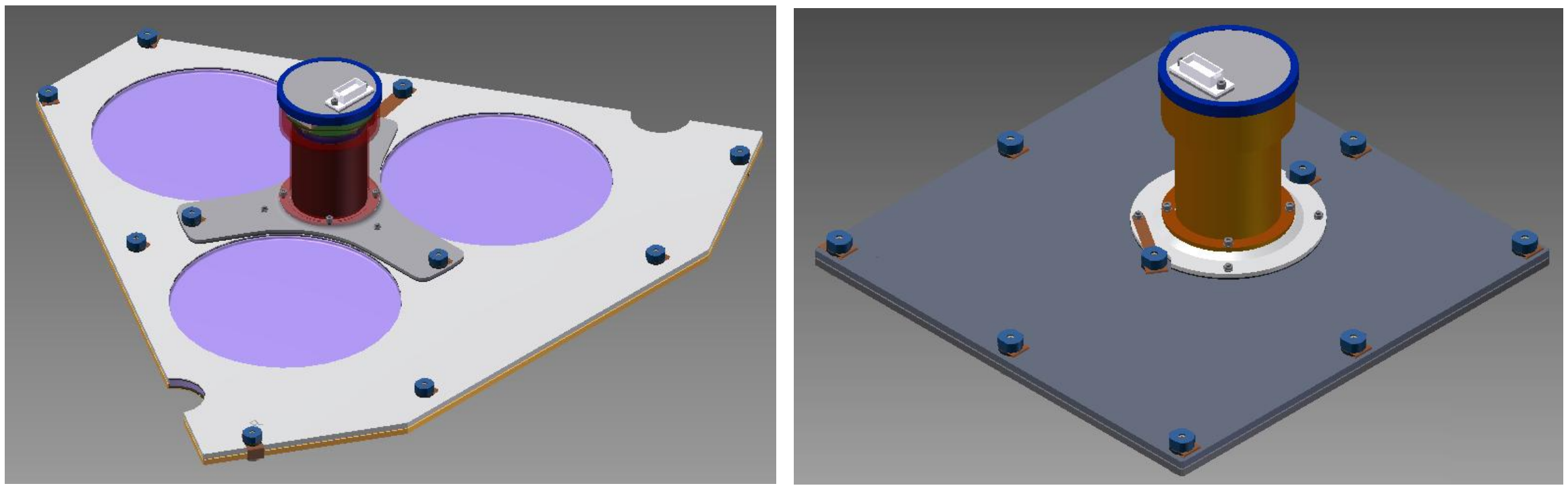

Figure 4 : One top ACD and one lateral ACD

## 1.2 Medium Energy X-ray telescope

ME consists of 3 detector boxes as shown in Figure 5. Each box has 576 Si-Pin detector pixels read out by 18 ASIC (Application Specified Integrated Circuit). Each ASIC is responsible for the readout of 32 pixels. The working temperature of Si-Pin detectors in orbit is from -50℃ to -5℃.

For each detector box, the collimators of ME confine 15 ASICs as narrow FOV(1° x 4°), 2 ASIC as wide FOV(4° x 4°) and one blind FOV. The layout of the FOVs in one detector box is also shown in Figure 5. Two in-orbit calibration radioactive sources ($^{241}$Am) are installed in small FOVs.

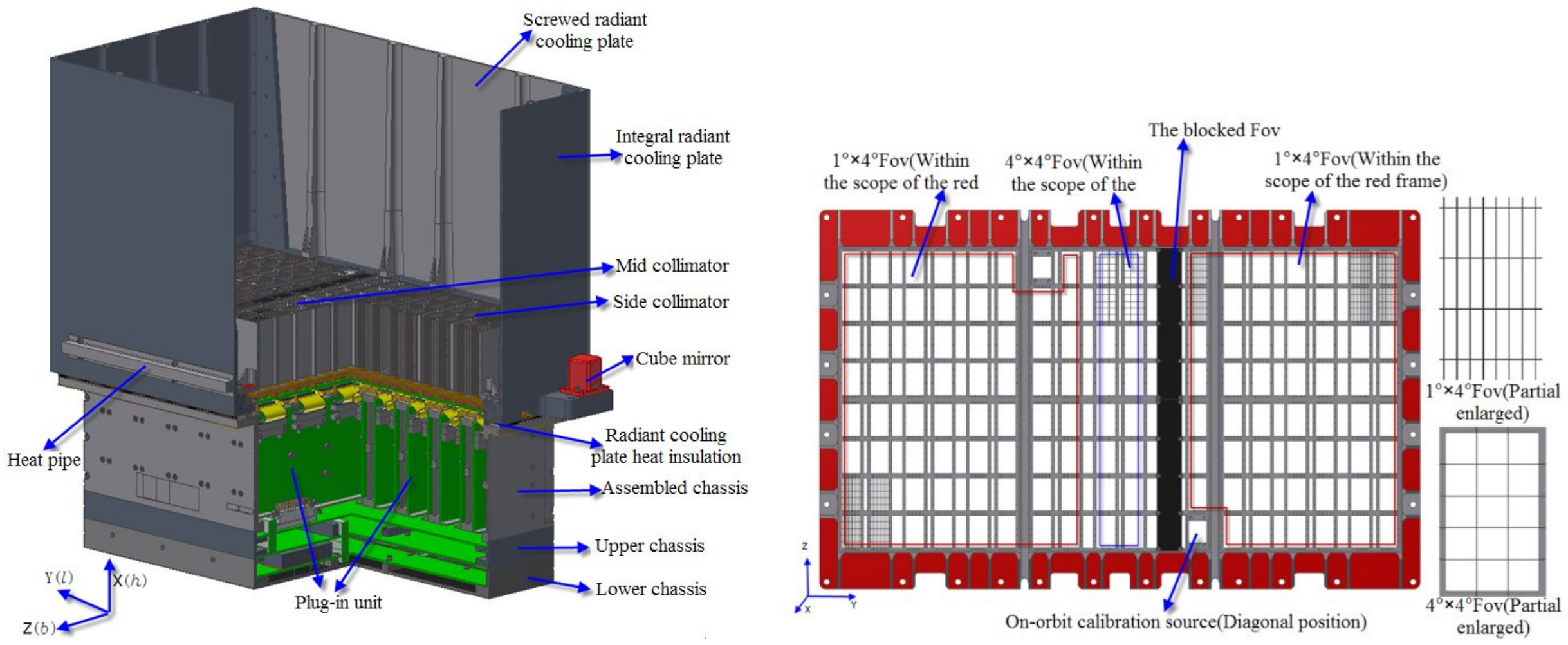

Figure 5 : Left: One detector box of ME. Right: Layout of FOVs in one detector box.

Si-Pin detectors are fixed on the ceramic chip by silver glue. When the energy of incident X-rays is greater than 25.5 keV (K-edge of Ag), they have some probability of penetrating the Si-Pin and react with silver. Ag emission lines will be generated due to the photoelectric effect with electrons in K-shell of Ag and detected by the Si-Pin detectors. Figure 6 showed that emission lines of Ag appeared in the detected energy spectrum when incident X-rays had energy more than 25.5 keV in the ground calibration experiments.

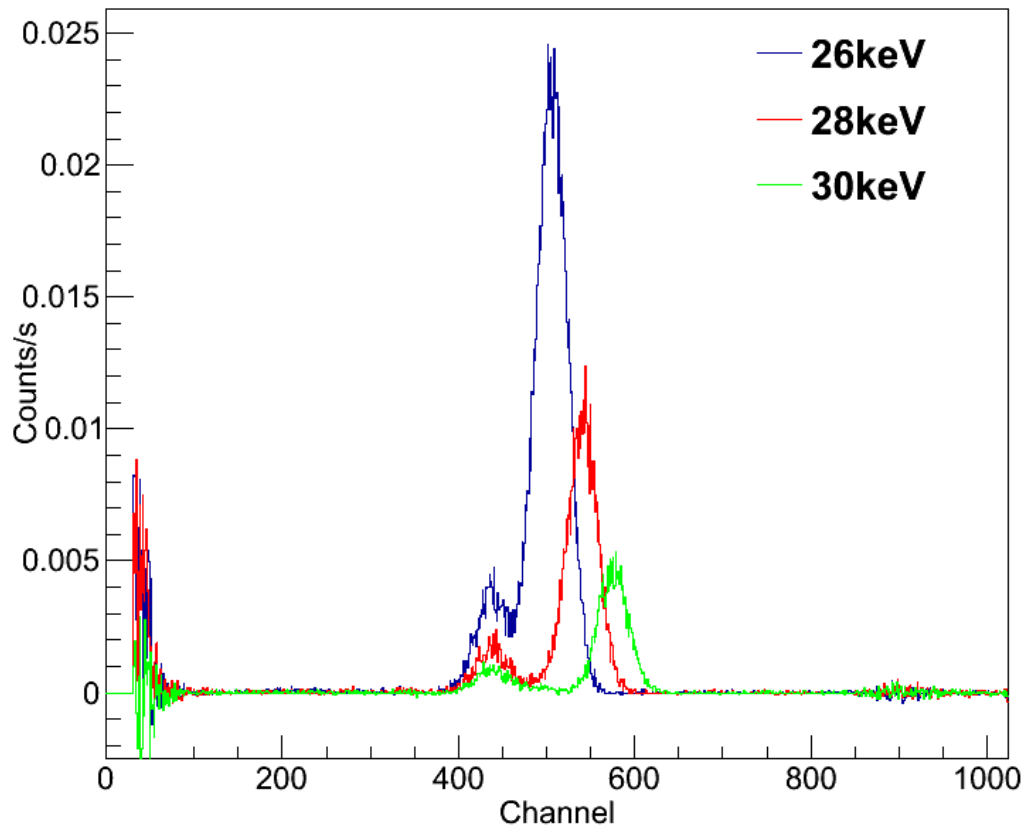

Figure 6 : The emission lines of Ag generated in the detected spectrum of Si-Pin detectors when the incident X-rays had energy more than 25.5 keV(K-edge of Ag)

### 1.3 Low Energy X-ray telescope

LE also consists of three detector boxes and each box contains 32 CCD236 which is a kind of Swept Charge Devices (SCD). CCD236 is a second-generation SCD, which has been developed by e2v company with full considerations of the requirement of LE. It is built with a sensitive area of about 4 $cm^2$ and has four quadrants. In each quadrant, the L-shaped electrodes guide the charge towards the diagonal first and then to a common readout amplifier in the central region. In the continuous readout mode, the total readout time is only about 1 ms with a guaranteed energy resolution. Figure 7 shows the detector box of LE and photograph of four pieces of CCD236. The working temperature in-orbit for CCD236 is about from -80℃ ~ -30℃.

For each detector box of LE, collimators define four kinds of FOVs. Twenty CCD236 have narrow FOVs with 1.6° x 6° . Six CCD236 have wide FOVs with 4° x 6° . Two CCD236 have blind FOVs and one of them has carried a $^{55}Fe$ radioactive source. Four CCD236 have a very large FOV with about 50~60° x 2~6° .

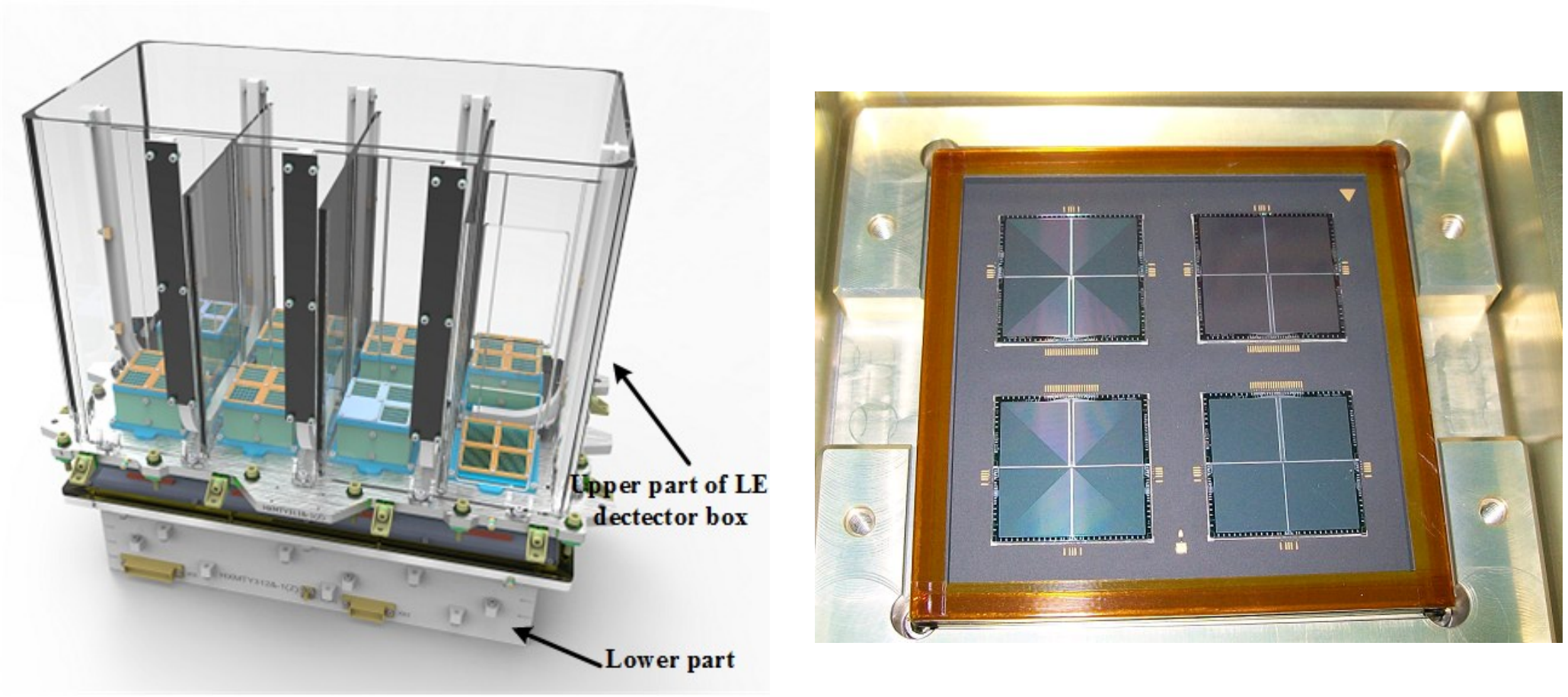

Figure 7: Left: One detector box of LE. Right: four pieces of CCD236

# 2. ENERGY SCALE AND RESOLUTION CALIBRATION

## 2.1 HE

Figure 8 shows the detected background spectrum from seven hours of blank sky. The background is dominated by internal activation effects. Prominent background lines due to activation of iodine by cosmic and SAA protons are seen at 31, 56, 67 and 191 keV. These four lines can be used to calibrate the Energy-Channel (EC) relation or energy scale. Background line of 191 keV is also used to monitor the long-term stability of EC relation in-orbit as shown in Figure 9.

The energy resolution for 31keV is used to estimate the energy resolution in orbit.

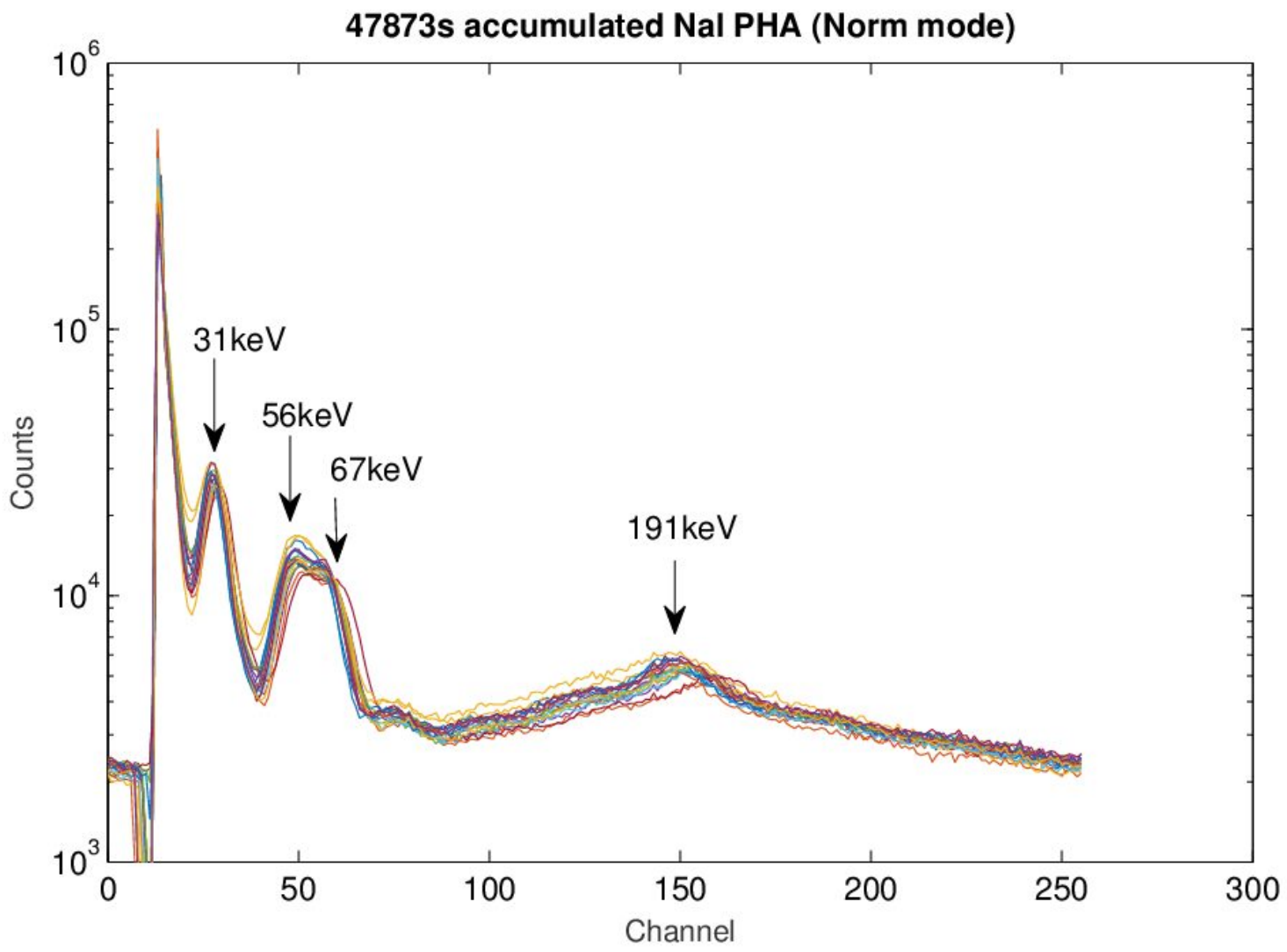

Figure 8: The observed background spectrum detected by 18 PHOSWISH detectors of HE. Background lines due to activation of iodine by cosmic and SAA protons are evident at 31, 56, 67, and 191 keV.

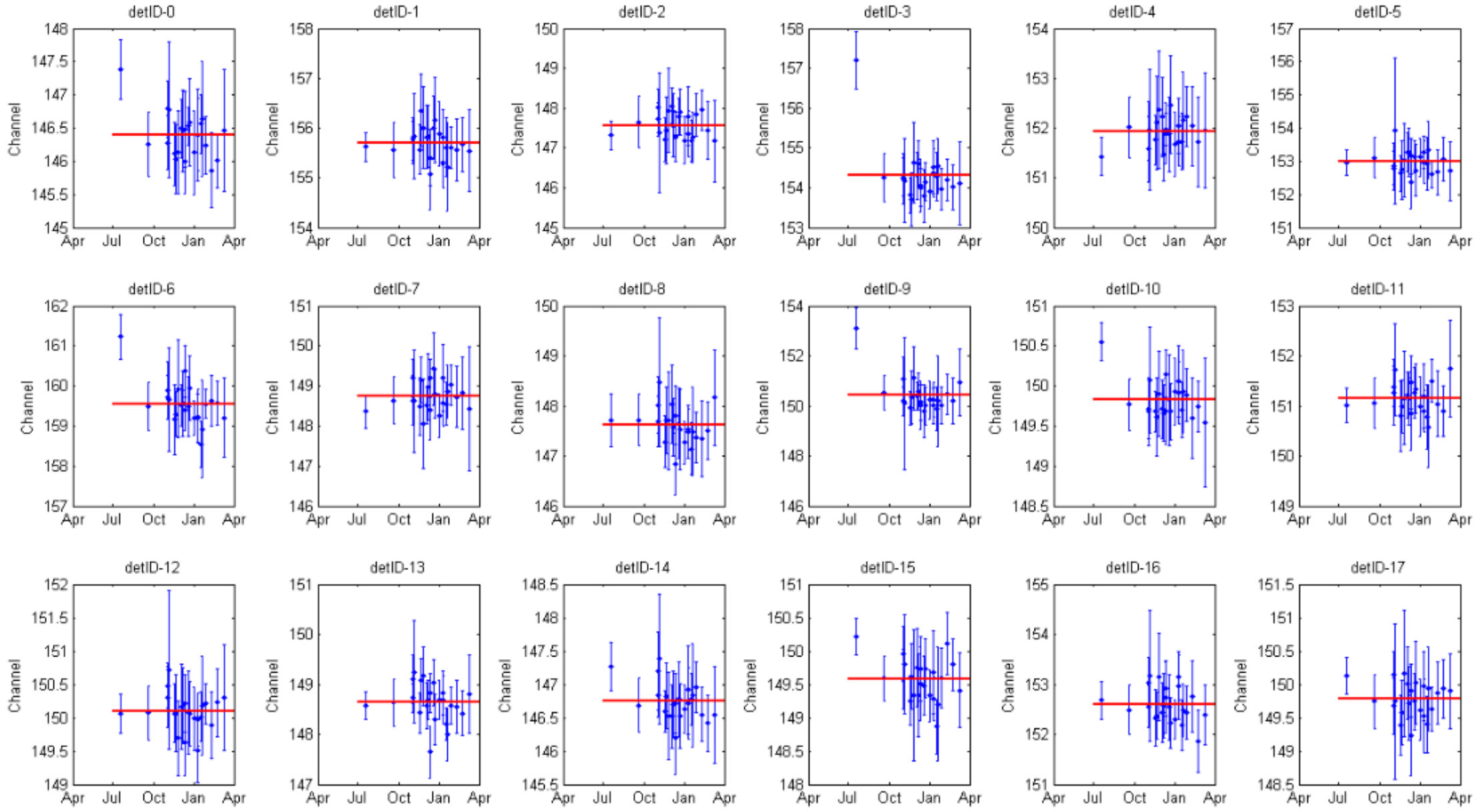

Figure 9: The peak of 191 keV background line can be used to monitor the stability of energy scale for HE detectors.

## 2.2 ME

The working temperature of ME in-orbit changed between -50℃ and -5℃. In the calibration experiment on ground, we found the Energy-Channel (EC) relation is linear from 11 keV to 30 keV using the spectrum of $^{241}$Am source. The slopes and intercepts of E-C relation of all pixels are not a constant at different temperatures as shown in Figure 10. The slopes and intercepts will be achieved from the linear interpolation at two adjacent temperatures for each pixel. In-orbit, the pixels carried $^{241}$Am and other pixels which have Ag line are used to verify the change of EC relation on ground. All blank sky data are analyzed to get the peak of Ag line. We found that EC on ground is also suitable. As shown in Figure 11, the mean value of Ag peak is equaled to 22.5 keV, the same with the expected value.

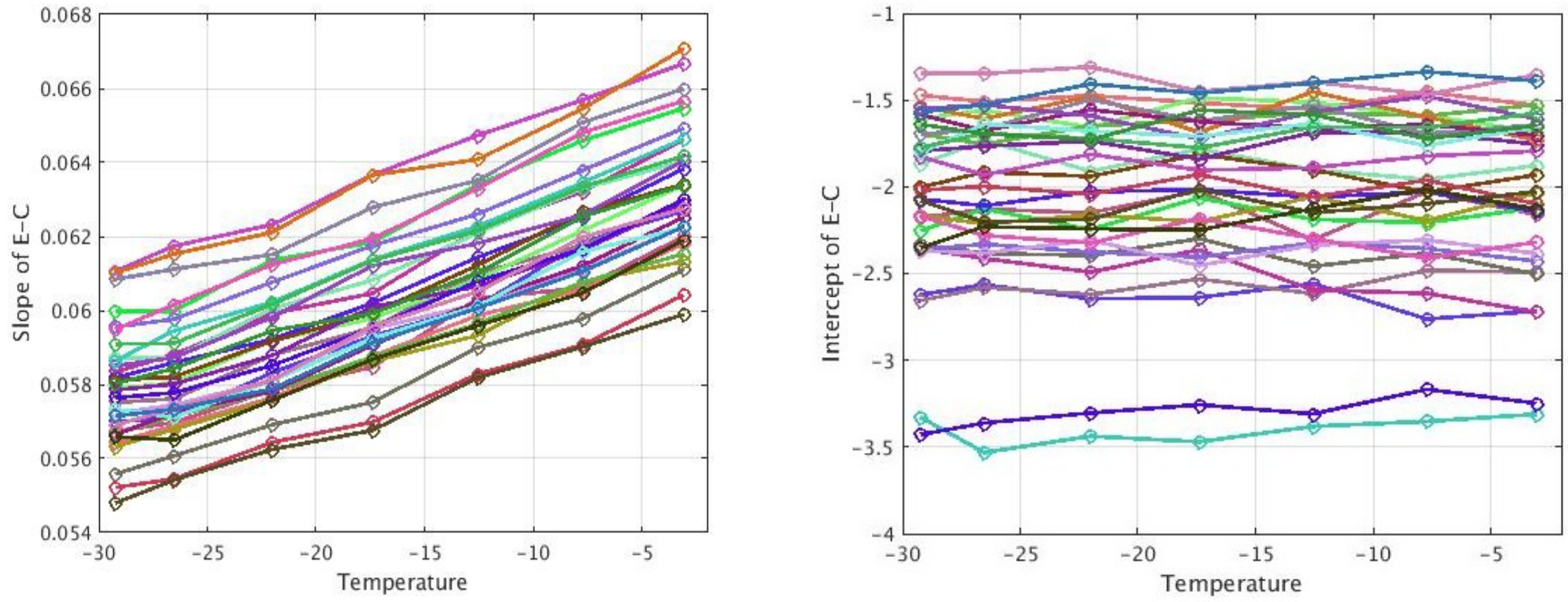

Figure 10: The slope and intercept of Energy-Channel relation for 32 pixels of one ASIC at different temperatures.

The pixels carried $^{241}$Am in orbit can also be used to estimate the variation of FWHM. We found that change is very small.

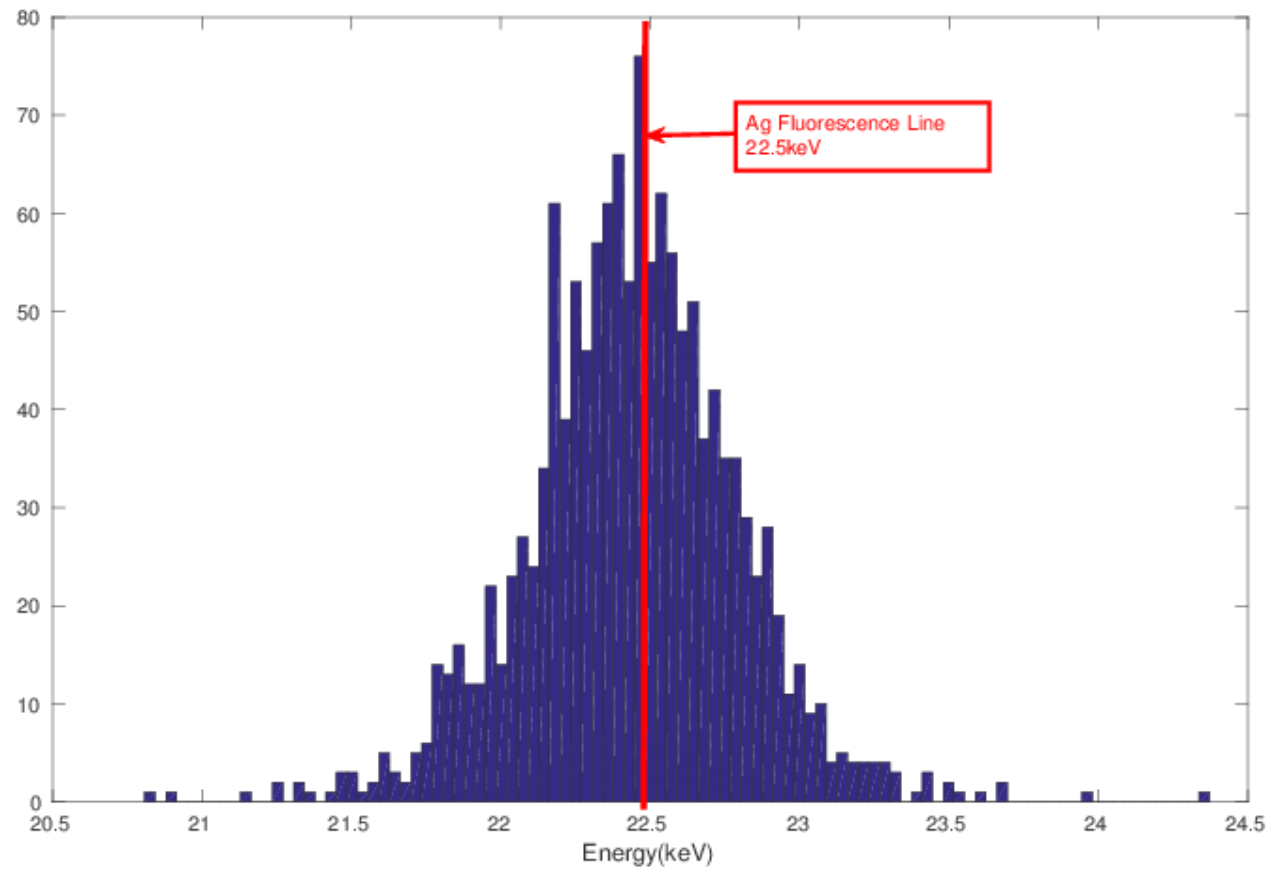

Figure 11: The distribution of Ag line peak for about 1200 pixels

### 2.3 LE

Ground calibration experiments showed that the linearity of LE was very good, and the slopes and intercepts also differed at different temperatures. The same method for ME was used to obtain the slopes and intercepts at different temperatures.

In-orbit, Cas A was used to verify the change of EC relation. Here, we used the observed spectrum of Chandra to get the model of Cas A and fitted the measured spectrum of LE to check the shape of residual distribution of two instruments. If the shape of the residual distribution is same of the two instruments, we think that it comes from the systematic error of the source model. Otherwise, the EC of LE has changed compared with the result on ground. From the fit result in Figure 12, the EC relation from 1.8 keV to 4 keV has changed, the peaks in this energy range are used to update the EC relation in-orbit.

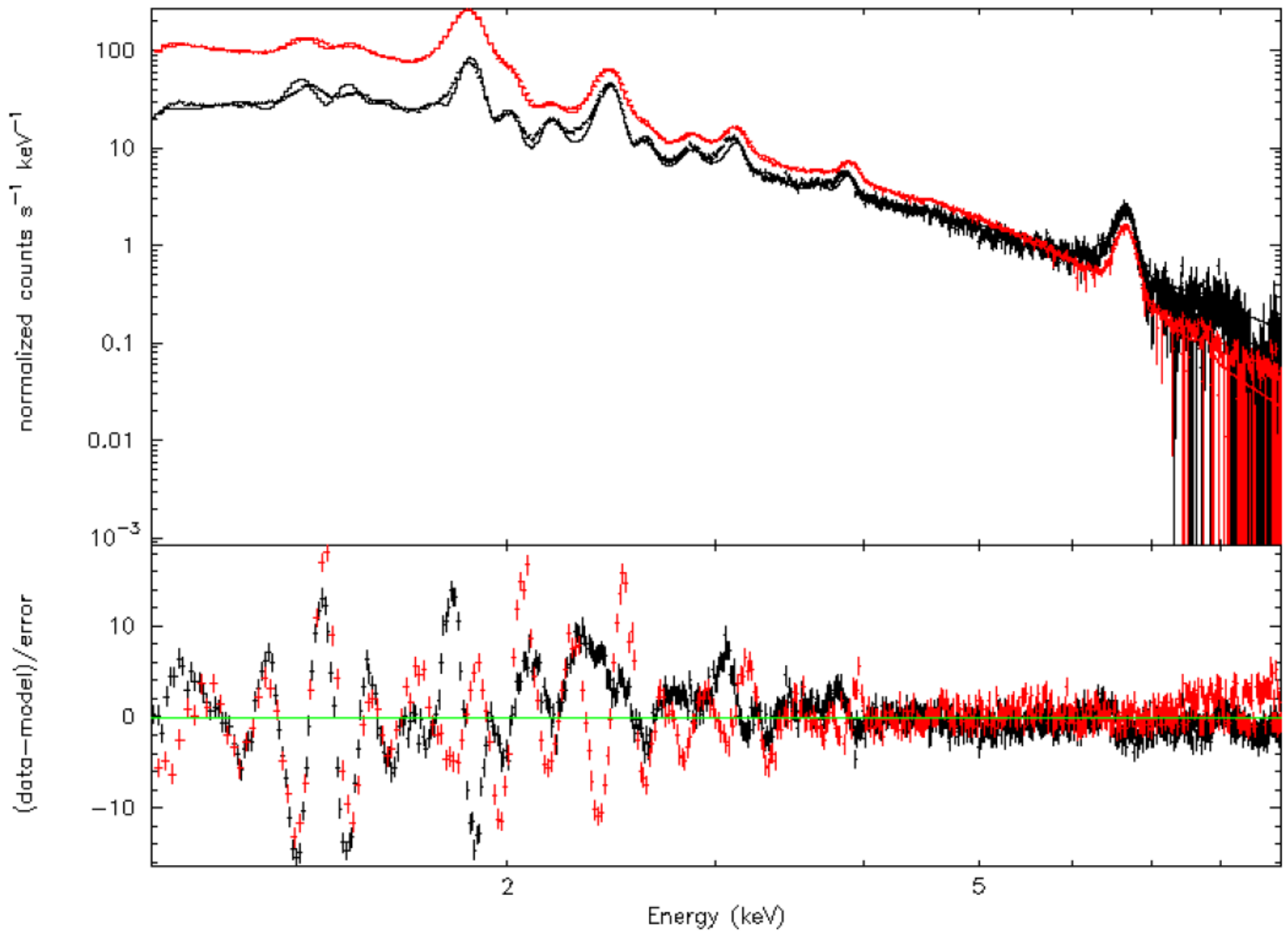

Figure 12: The red dot is the Cas A spectrum measured by Chandra. The black dot is the Cas A spectrum measured by LE. The EC relation from 1.8 keV to 4 keV was different with the on-ground results from the shape distribution of residual distribution.

# 3. EFFECTIVE AREA CALIBRATION

## 3.1 The Crab Pulsar as a Calibration Source

The Crab Nebula, with its pulsar has been served as a primary calibration target for many hard X-ray instruments because of its brightness, relative stability, and simple power-law spectrum. The Crab is too bright for most CCD based focusing X-ray instruments because of the pileup effect, but for LE, there are no problems of pileup.

As a collimated telescope, Insight-HXMT has to construct its background model to estimate the background level. There is no on/off observation and it is relatively hard for Insight-HXMT to estimate background because the particle-induced background varies significantly with time and orbit. In order to avoid the influence of background and independently obtain the systematic error of calibration, Crab Pulsar is used to calibrate the effective area of three payloads.

Figure 13 shows the pulse profiles of the Crab pulsar for the entire range of three payloads. The spectrum was generated with the whole pulse and the spectrum of the background is obtained from the phase between 0.6 and 0.8. To acquire the spectral parameter, the spectrum of the Crab pulsar observed by RXTE/PCA and RXTE/HEXTE in the year of 2011(Reference) was calculated with the same method. As shown in Figure 14, the model of the pulsed spectrum can be fitted by LOGPAR(Reference) according to the joint fitting result of two instruments. The fitting parameter alpha is equal to 1.52, beta is equal to 0.139, and norm is equal to 0.448.

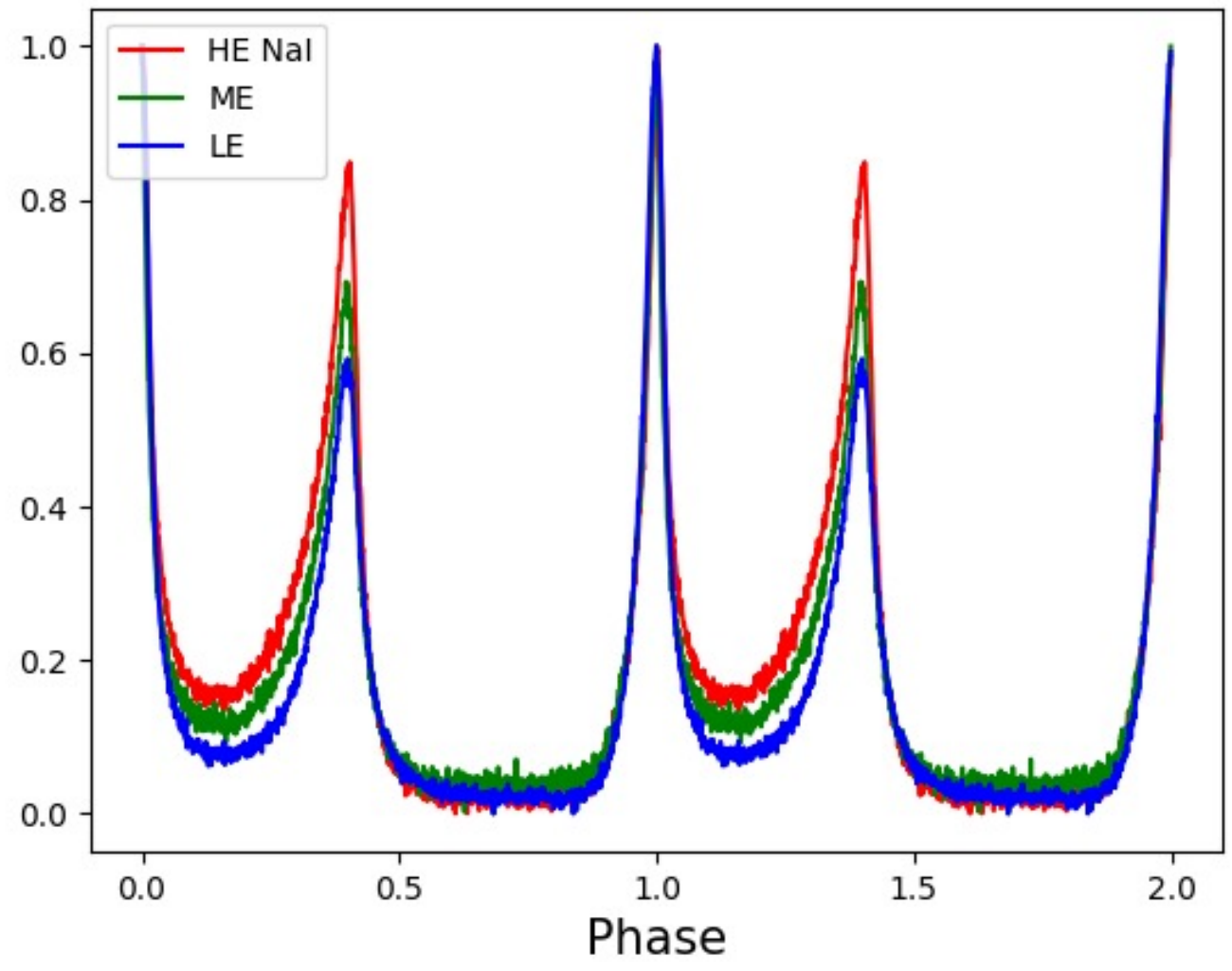

Figure 13: X-ray profiles of Crab Pulsar detected by three payloads of Insight-HXMT. Phase 0 represents the position of the main radio peak. The un-pulsed component is subtracted from phase 0.6 to phase 0.8.

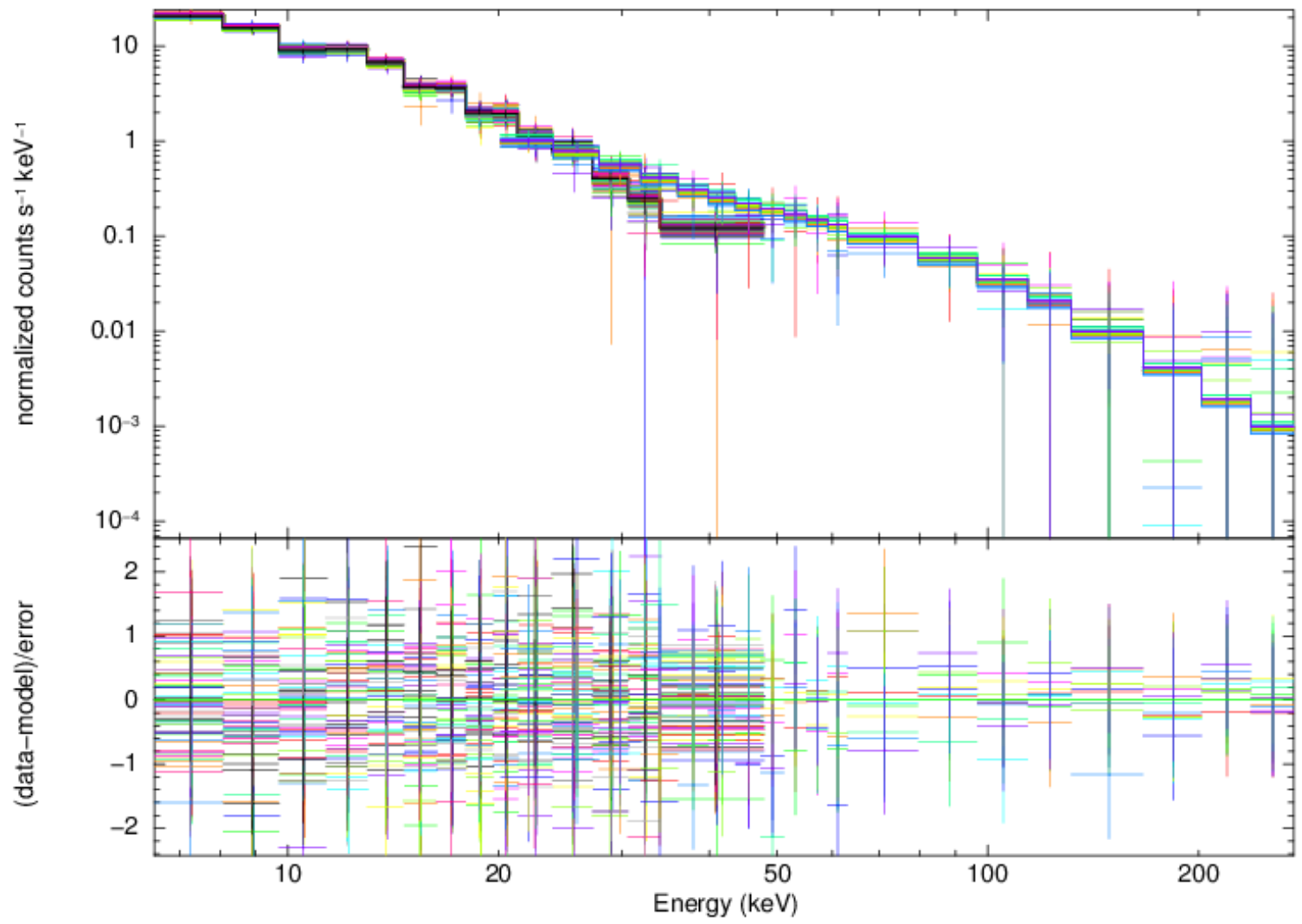

Figure 14: The Crab Pulsar spectrum observed by RXTE/PCA and HXTE in the year of 2011. They jointly determined the model of Crab Pulsar as LOGPAR.

### 3.2 Results of in-orbit effective areas

Pure Monte Carlo model of effective area is very difficult to fit the pulsar spectrum well because the absorption and scattering of anticoincidence detectors, non-uniform response of the detectors and so on. Finally, it was decided to use an empirical function to modify the simulated effective areas. The comparisons of the simulated and modified effective areas are shown for three payloads are shown in Figure 15. With the new modified effective areas and fixed the parameters of Crab pulsar, the residual distributions of Crab Pulsar spectrum observed by the three payloads can be found in Figure 16. There is no additional systematics added, 2% will be sufficient for a good fit.

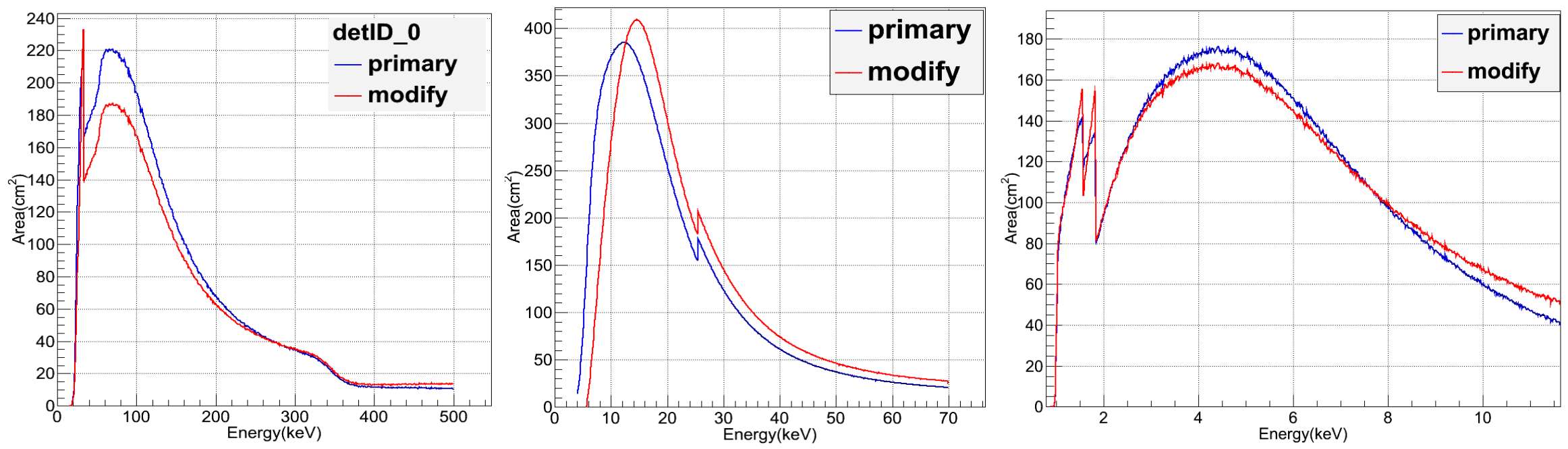

Figure 15: The comparisons of the primary simulated and modified effective areas. The left one is one NaI(Tl) detector, the middle one is small FOV pixels of ME, and the right one is small FOV CCD236 of LE.

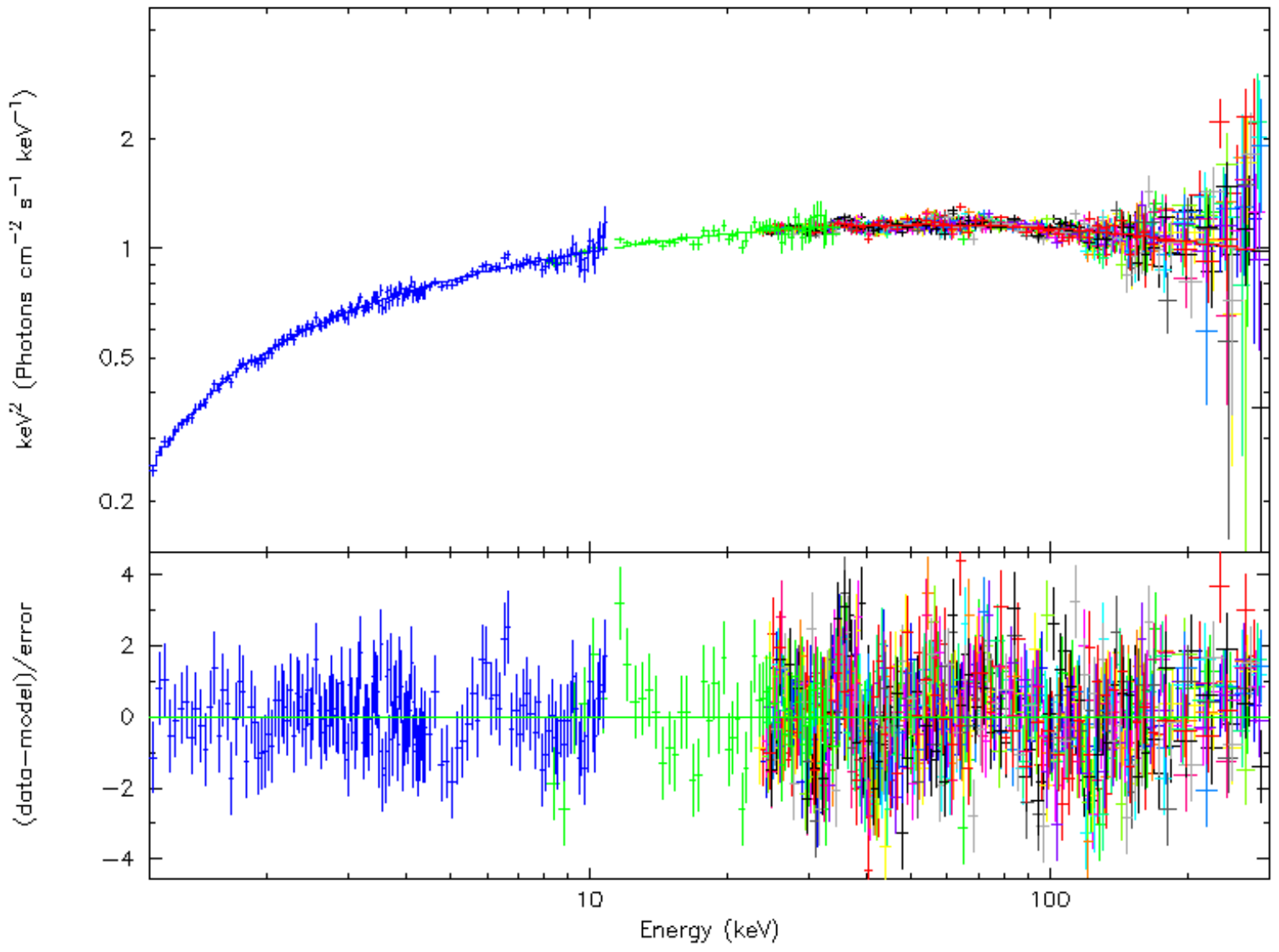

Figure 16: The residual distributions of Crab Pulsar spectrum observed by three payloads of Insight-HXMT with modified effective areas and fixed parameters of Crab Pulsar.

# 4. TIMING CALIBRATION

The timing information of Insight was verified through observations of the Crab Pulsar based on the coordinated radio and Fermi-LAT observations. Figure 17 shows the timing residuals of the Pulsar for Insight-HXMT, Radio Telescopes in Xinjiang and Yunnan and Fermi-LAT by TEMPO2. The timing residuals observed from Insight-HXMT are almost the same with the other two instruments. At the end, absolute time accuracy of HXMT is better than 100us from the root-mean-square of timing residual 51 us.

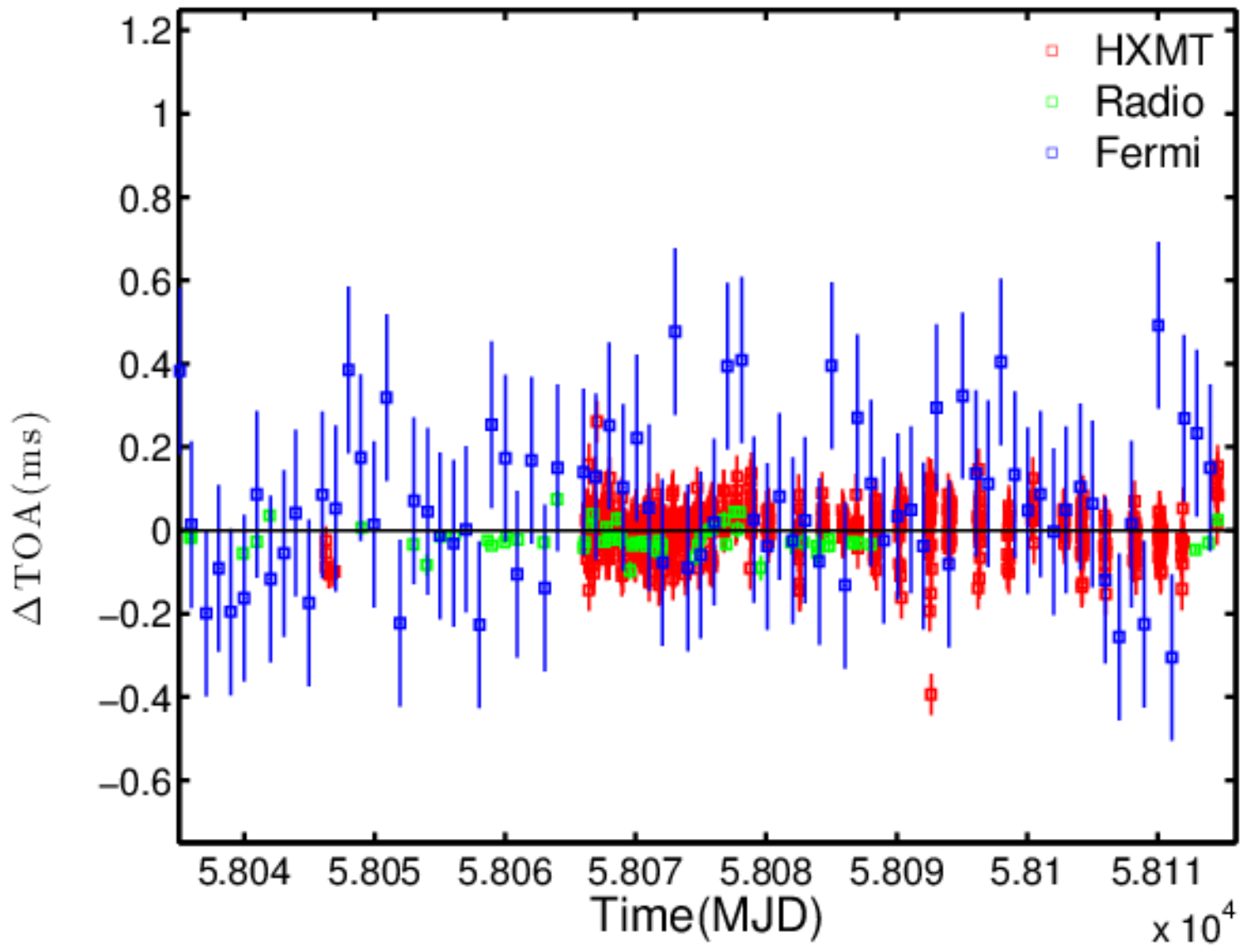

Figure 17: The residuals of TOA of Crab for Radio, Fermi/LAT, and HXMT. The absolute time accuracy of HXMT is better than 100us.

## 5. SUMMARY

Insight-HXMT has been in-orbit for nearly one year. It has performed well and operated within requirements. The primary calibration results have been tested and used in the science analysis. The long-term monitor of performances has been carried out routinely.

## ACKNOWLEDGEMENTS

This work is supported by the National Key Research and Development Program of China (2016YFA0400800) and the National Natural Science Foundation of China under grants 11503027. This work made use of the data from the Insgiht-HXMT mission, a project funded by China National Space Administration (CNSA) and the Chinese Academy of Sciences (CAS).

## REFERENCES

[1] Li T.P. & Wu M., "A Direct Restoration Method for Spectral and Image Analysis", Astrophysics and Space Science 206, 91-102 (1993)
[2] Zhou X., Li X.Q., Xie Y.N., Liu C.Z., Zhang S., Wu J.J., Zhang J., Li X.F., Zhang Y.F., Li B., Hu H.L., Chen Y.P., Jiang W., & Li Z.S., "Introduction to a calibration facility for hard X-ray detectors", Experimental Astronomy,38,433-441 (2014)
[3] Zhang S., Chen Y.P., et al., "The x-ray facilities in-building for calibrations of HXMT", Proc. SPIE 9144, 55(2014)
[4] Li T.P., et al., "Insight-HXMT observations of the first binary neutron star merger GW170817", Science China 61(3): 031011-031018 (2018)
[5] Ge M.Y., Lu F.J., et al., "X-ray phase-resolved spectroscopy of PSRs B0543+21, B1509-58, and B0540-69 with RXTE", The Astrophysical Journal Supplement Series, 199:32-52 (2012)
[6] Wang, N., Wu, M., Manchester, R. N., Zhang, J., et al., "Pulsar timing at Urumqi Astronomical Observatory: observing system and results", Mon. Not. R. Astron. Soc. 328, 855–866 (2001).
[7] Luo, J. T., Chen, L., Han, J. L., Esamdin, A., Wu, Y. J., Li, Z. X., Hao, L. F., Zhang, X. Z., A, "digital pulsar backend based on FPGA". RAA 17, 9 (2017)
[8] Hobbs, G. B., Edwards, R. T. & Manchester, R. N. "TEMPO2, a new pulsar-timing package -I. An overview". Mon. Not. R. Astron. Soc. 369, 655–672 (2006)
[9] Abdo, A. A., et al. "Fermi Large Area Telescope Observations of the Crab Pulsar And Nebula", Astrophys. J. 708, 1254–1267 (2010)
[10] Yang X.J., Lu F.J., and Chen L., "High Spatial Resolution X-Ray Spectroscopy of Cas A with Chandra", Chinese Journal of Astronomy and Astrophysics, 8, 439-450(2008)